# A Review of Dispersion Control Charts for Multivariate Individual Observations

Jimoh Olawale Ajadi[1], Zezhong Wang[2], and Inez Maria Zwetsloot[3]

Department of Systems Engineering and Engineering Management, City University of Hong Kong, Kowloon, Hong Kong

20th December 2019

**Abstract**

A multivariate control chart is designed to monitor process parameters of multiple correlated quality characteristics. Often data on multivariate processes are collected as individual observations, i.e. as vectors one at the time. Various control charts have been proposed in the literature to monitor the covariance matrix of a process when individual observations are collected. In this study, we review this literature; we find 30 relevant articles from the period 1987-2019. We group the articles into five categories. We observe that less research has been done on CUSUM, high-dimensional and non-parametric type control charts for monitoring the process covariance matrix. We describe each proposed method, state their advantages, and limitations. Finally, we give suggestions for future research.

___________________________________

[1] Mr. Jimoh Olawale Ajadi is a PhD student in the Department of Systems Engineering and Engineering Management at City University of Hong Kong. His email address is joajadi2-c@my.cityu.edu.hk

[2] Miss. Zezhong Wang is a PhD student in the Department of Systems Engineering and Engineering Management at City University of Hong Kong. Her email address is zezhowang3-c@my.cityu.edu.hk

[3] Dr. Inez Maria Zwetsloot is an assistant professor in the Department of Systems Engineering and Engineering Management at City University of Hong Kong. Her email address is i.m.zwetsloot@cityu.edu.hk

## 1.    Introduction

Multivariate dispersion control charts monitor the process variability of multiple correlated quality characteristics. The most common method for monitoring the covariance matrix of the process, is the generalized variance chart proposed by Alt (1985). This chart applies the determinant of the estimated covariance matrix as the monitoring statistic. Over the past decades, many methods have been developed for monitoring the covariance matrix. To obtain an estimate of the covariance matrix, usually a sample of $n > p$ observations are collected. Here $p$ is the number of correlated quality characteristics to be monitored. We refer to this as *grouped data*. For details on multivariate dispersion charts for grouped data, see the review articles by Yeh, Lin, and McGrath (2006) and Bersimis, Psarakis, and Panaretos (2006).

Observations are not always available in groups. When data are obtained vector-by-vector, we refer to this as *individual observations*. The challenge now becomes how to obtain an estimate of the variance-covariance matrix. The first method for monitoring the dispersion using individual observations was proposed by Healy (1987). Many other methods are also available. Recently, Ajadi and Zwetsloot (2019) compared the performance of selected multivariate dispersion charts based on individual and grouped observations. The authors concluded that monitoring methods based on individual observations are quicker in detecting sustained shifts in the process variability. This conclusion motivates the current study. The objective of this article is to review the literature on control charts for monitoring the dispersion of multivariate processes where data is collected as individual vectors (observations).

With the exception of Yeh, Lin, and McGrath (2006), we have not seen a comprehensive literature review on multivariate dispersion charts for individual observations. Since 2006, many paper (25) have been written on this topic. In this review, we also include the articles that were reviewed by Yeh, Lin, and McGrath (2006). This brings the total number of reviewed articles to 30, published in the period between 1987 and 2019.

We classify the existing methods for monitoring the process variability into five categories based on the type of chart. The first category is the CUSUM-type dispersion charts where we discuss 4 articles. In the second group, the MEWMA-type dispersion charts, we review 14 articles. Next, we discuss 6 articles in the Shewhart-type category. In the fourth and fifth category, we review the non-parametric control charts (2 articles) and the multivariate charts with high dimensions (4 articles), respectively.

We have excluded papers discussing multivariate charts like the Hotelling $T^2$ (see Hotelling (1947)) and MEWMA (see Lowry et al. (1992)) which react to the shifts in the process variability but that are specially designed to monitor the process mean vector. Likewise, the charts proposed by Zhang and Chang (2008) are not specifically designed for monitoring the covariance matrix. The focus of the current study is to review charts that are especially designed to monitor the covariance matrix of the process.

We give an overview of the literature included in this study in Section 2; we state the model and assumptions underlying the reviewed methods in Section 3. In Sections 4 through 8, we discuss the literature for each of the 5 types of monitoring methods. Section 9 provides conclusions and future research directions.

## 2. Overview of the Literature

We studied more than 200 research articles published from 1987 to 2019 on multivariate control charts. Google scholar was employed as the database for searching relevant articles. Thereafter, we determined the articles to be included in our study based on the following selection criteria- that is, articles that:

- focus on monitoring the process covariance matrix,
- are designed for multivariate individual observations,
- focus on Phase II monitoring applications,
- employ control charts as monitoring tool.

Consequently, we identified thirty articles. Detailed information of the selected articles is displayed in Figures 1&2 and Tables 1&2.

The first dispersion chart for individual observations was published in 1987, when Healy (1987) proposed CUSUM-type dispersion charts. Later in 2001, Chan and Zhang (2001) introduced another CUSUM dispersion chart. In Figure 1, we notice that 21 articles were published between 2010 and 2019, and 10 of them are MEWMA-type dispersion charts. Four articles are published each in 2013, 2017, and 2019; it is the largest number in history. We observe that less research has been done on CUSUM (4), high-dimensional (4) and the non-parametric (2) charts for monitoring the process covariance matrix based on individual observations.

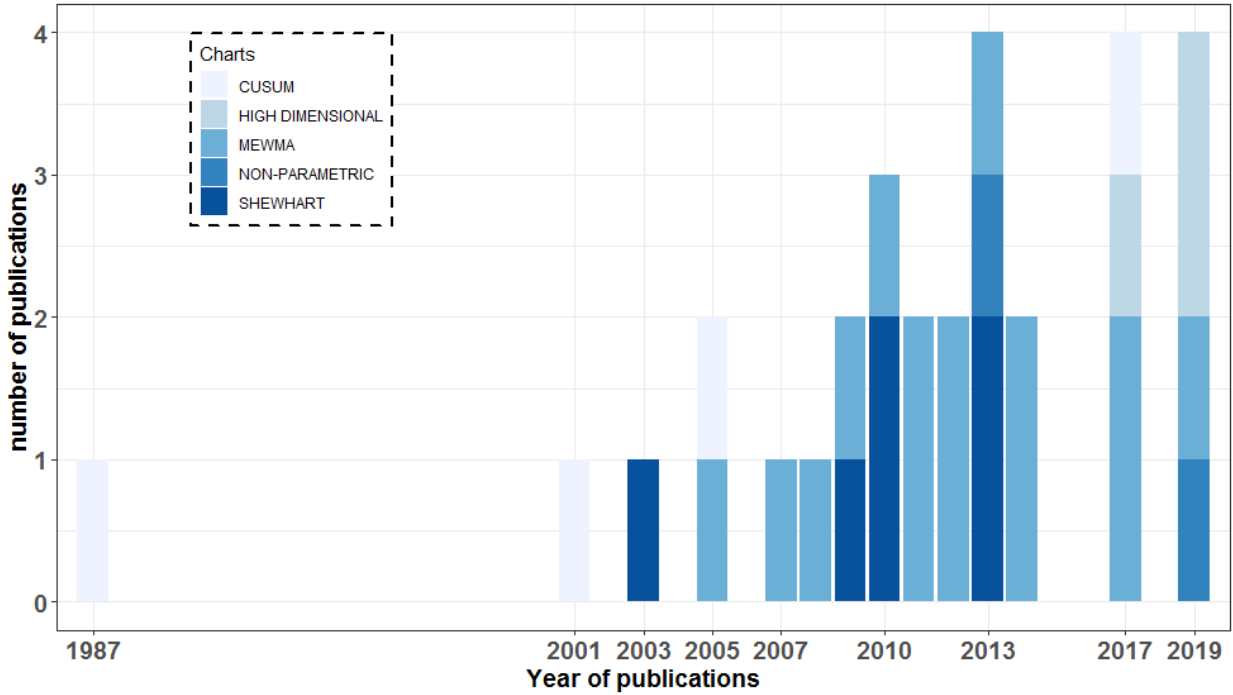

Figure 1: Number of articles per year on multivariate dispersion chart based on individual observations.

Figure 2 displays the number of publications sorted by journal; eight journals that each have only one publication are grouped as "Others". We notice that Quality and Reliability Engineering International (5 publications) has the highest number of published articles; followed by International Journal of Production Research, Computational Statistics and Data Analysis and Journal Quality Technology, each have three published articles.

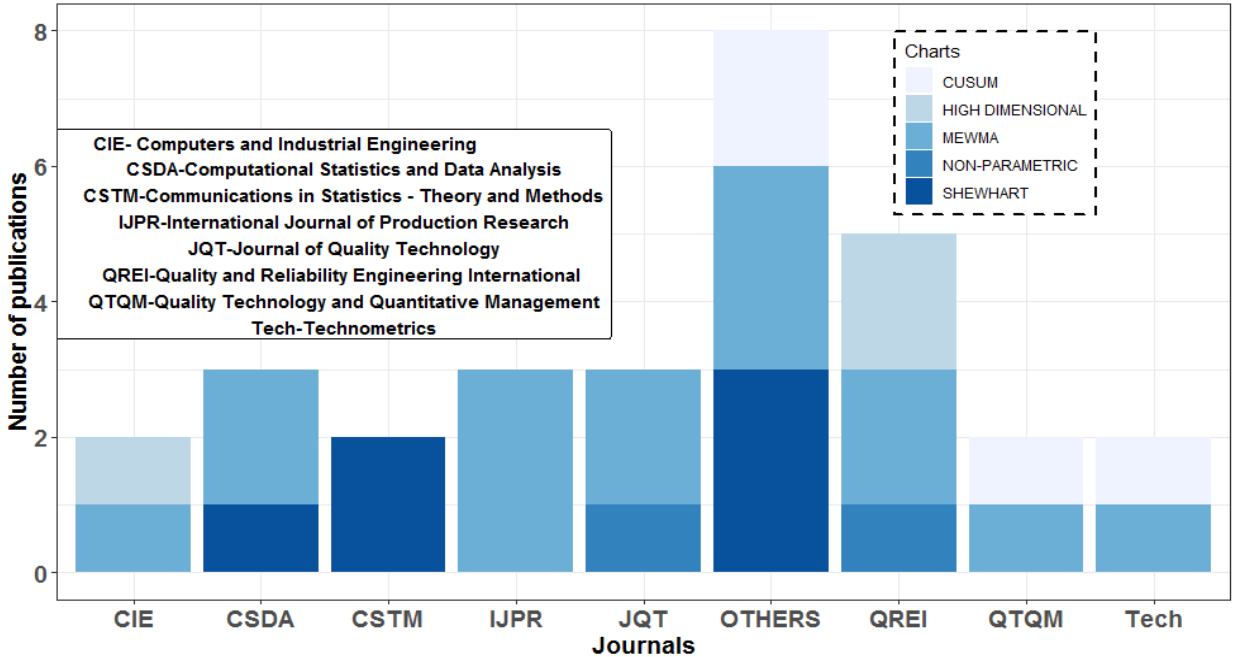

Figure 2: Number of publications that are based on multivariate dispersion charts for the individual observations by journal.

We display the fifteen articles with the highest number of citations in Table 1. The article with the highest number of citations is Healy (1987). Among the recent articles (2017-2019), though not displayed in Table 1, Gunaratne et al. (2017) has the most citations (6). This article discusses a powerful algorithm to improve the performance of the charts proposed by Huwang, Yeh, and Wu (2007) in detecting changes in the process covariance matrix for high dimensional data.

| | Authors | Journal | Number of Citations | Year of Publications |
|---|---|---|---|---|
| 1 | Healy | Technometrics | 388 | 1987 |
| 2 | Huwang, Yeh, and Wu | Journal of Quality Technology | 98 | 2007 |
| 3 | Hawkins and Maboudou-tchao | Technometrics | 91 | 2008 |
| 4 | Yeh, Huwang, and WU | IIE Transactions | 80 | 2005 |
| 5 | Zhang, Li, and Wang | Computational Statistics and Data Analysis | 62 | 2010 |
| 6 | Chan and Zhang | Statistical Sinica | 53 | 2001 |
| 7 | Mason, Chou, and Young | Communications in Statistics - Theory and Methods | 41 | 2009 |
| 8 | Khoo and Quah | Quality Engineering | 36 | 2003 |
| 9 | Cheng and Thaga | Quality Technology and Quantitative Management | 30 | 2005 |
| 10 | Maboudou-Tchao and Hawkins | Journal of Quality Technology | 28 | 2011 |
| 11 | Li et al. | Journal of Quality Technology | 28 | 2013 |
| 12 | Shen, Tsung, and Zou | International Journal of Production Research | 25 | 2014 |
| 13 | Yeh, Li, and Wang | International Journal of Production Research | 21 | 2012 |
| 14 | Memar and Niaki | Quality and Reliability Engineering International | 20 | 2009 |
| 15 | Maboudou-Tchao and Agboto | Computational Statistics and Data Analysis | 19 | 2013 |

Table 1: The 15 most often cited articles on multivariate dispersion charts for individual observations (citation count from google scholar as of 28 October 2019).

Table 2 provides an overview of selected articles grouped by the type of chart and the employed monitoring statistics. Eight articles use the trace of the estimated covariance matrix to compute monitoring statistics. Other methods such as the Wilk's statistic, the Alt's likelihood ratio, and the $L_1$ −norm and $L_2$ −norm, are applied by other papers.

| Type of Chart | Monitoring Statistic | Reference |
|---|---|---|
| Shewhart | Wilk's statistic | Mason, Chou, and Young (2009, 2010) Djauhari (2010) |
| | Chi-squared | Khoo and Quah (2003) |
|  | Alt's likelihood ratio | Maboudou-Tchao and Agboto (2013) |
|  | Trace | Hwang (2017) |
| MEWMA | Norm statistic | Yeh, Huwang, and Chien-Wei (2005) |
|  |  | Memar and Niaki (2009) |
|  |  | Shen, Tsung, and Zou (2014) |
|  |  | Fan et al. (2017) |
|  | Trace | Huwang, Yeh, and Wu (2007) |
|  |  | Memar and Niaki (2011) |
|  |  | Tsai et al. (2012) |
|  |  | Gunaratne et al. (2017) |
|  |  | Alfaro and Ortega (2019) |
|  | Alt's likelihood ratio | Hawkins and Maboudou-tchao (2008) |
|  |  | Zhang, Li, and Wang (2010) |
|  |  | Maboudou-Tchao and Hawkins (2011) |
|  |  | Maboudou-Tchao and Diawara (2013) |
|  |  | Wang, Yeh, and Li (2014) |
|  | Sum of all elements | Memar and Niaki (2011) |
| MCUSUM | Maximum of CUSUM statistics of both process mean vector and dispersion | Cheng and Thaga (2005) |
| | CUSUM statistic | Chan and Zhang (2001) |
| | | Bodnar and Schmid (2017) |
| Non-Parametric | Trace of EWMA statistic | Li et al. (2013) |
| | Spatial rank | Huwang, Lin, and Yu (2019) |
| High Dimensions | Dissimilarity index | Huwang et al. (2017) |
| | Trace | Gunaratne et al. (2017) |
| | | Kim et al. (2019) |
| | | Li and Tsung (2019) |

Table 2: Monitoring statistics for different types of charts.

## 3. Model and Assumptions

Most of the reviewed articles assume that the data follow a multivariate normal distribution, $i.e.$ $X_i \sim N_p(\boldsymbol{\mu}, \boldsymbol{\Sigma})$, where the process mean vector and covariance matrix of a $p$ correlated quality characteristics are denoted by $\boldsymbol{\mu}$ and $\boldsymbol{\Sigma}$, respectively.

In some of these articles, the parameters are estimated in Phase I, in this scenario, we assume that $X_j$ is observed in Phase I at times $j = 1,2,3,\ldots,m$ and $X_i$ is observed in Phase II at times $i = 1,2,3,\ldots$ each equidistant in time. Note that $m$ is the Phase I sample size. The process parameters of most of the charts we reviewed are assumed to be known. When the process is in-control, the in-control parameters are $\boldsymbol{\mu} = \boldsymbol{\mu_0}$ and $\boldsymbol{\Sigma} = \boldsymbol{\Sigma_0}$.

Often, $X_i$ is standardized by transforming it to $Y_i$ as

$$Y_i = \Sigma_0^{-\frac{1}{2}}(X_i - \mu_0) \qquad (1)$$

Thus, $Y_i$ follows a standardized multivariate normal distribution $N(\mu_Y, \Sigma_Y)$, where $\mu_Y = \Sigma_0^{-\frac{1}{2}}(\mu - \mu_0)$ and $\Sigma_Y = \Sigma_0^{-\frac{1}{2}} \Sigma \Sigma_0^{-\frac{1}{2}}$. When the process is in-control, $Y_i \sim N(0, I_p)$, where $I_p$ is a $p \times p$ identity matrix. Throughout the article, the operator $tr(.)$ and $|.|$ respectively denote the trace and determinant of the sample covariance matrix. Also, $||A||_1$ and $||A||_2$ denote the $L_1$-norm and the $L_2$-norm of matrix $A$ respectively.

## 4. Charts Based on CUSUM Statistics

In this section, we review four articles that propose CUSUM-type dispersion charts for individual observations. CUSUM control charts are an effective technique for detecting small process shifts. This type of chart is usually employed with individual observations (Montgomery (2009)). In this section, we review four articles that proposed multivariate CUSUM-type dispersion charts for individual observations. Healy (1987) and Cheng and Thaga (2005) used likelihood ratio to build charting statistics. Chan and Zhang (2001) developed CUSUM charts using a projection pursuit (PP) method. Bodnar and Schmid (2017) modified four CUSUM-type charts for monitoring the covariance matrix of multivariate time series data. All the charts proposed in these articles are designed to detect changes in normally distributed data. Next, we will briefly introduce these charts and their performances. We give information about each chart, such as the subgroup size (n), and $p$ simulated in each of the articles in Table 3.

| Authors | Chart name | n | p | Distribution | Process parameters in Phase I |
|---|---|---|---|---|---|
| Healy (1987) | CUSUM | NIL | NIL | normal | Known |

| Chan and Zhang (2001) | $MCD_1$ | n=1,2,5,10 | 2,3,4 | normal | Known |
| Cheng and Thaga (2005) | Max-CUSUM | n=1 | 2,5 | normal | Known |
| Bodnar and Schmid (2017) | Modified CUSUM | n=1 | 4 | normal and non-normal | Known |

Table 3: Overview of the CUSUM-type dispersion chart.

## *4.1 MCUSUM Charts Based on Likelihood Ratio*

Healy (1987) used CUSUM theory to derive a scheme for detecting proportional shifts in the variance-covariance matrix from $\mathbf{\Sigma_0}$ to $C\mathbf{\Sigma_0}$, where the correlation between two variables remains constant. The log likelihood ratio based CUSUM statistic $T_i$ is computed as

$$T_i = \max\left(T_{i-1} + \log\frac{f_B(X_i)}{f_G(X_i)}, 0\right) \qquad (2)$$

where $f_B$ and $f_G$ are the density functions corresponding to the in-control and out-of-control distribution of the data respectively. Under the normality assumption and a shift from $\mathbf{\Sigma_0}$ to $C\mathbf{\Sigma_0}$, Eq. [2] simplifies to

$$T_i = \max\left(T_{i-1} + (X_i - \mu_0)'\Sigma_0^{-1}(X_i - \mu_0) - K, 0\right)$$

where $K = p\log(C)\,[C/(C-1)]$. This method can be adapted to the case where the sample size is greater than 1.

Cheng and Thaga (2005) employed a maximum CUSUM statistic to extend this scheme and developed their own chart, which they called the Maximum Multivariate Cumulative Sum (Max-MCUSUM) control chart. The proposed method can monitor changes in the process mean vector and covariance matrix simultaneously. They compared the Max-MCUSUM chart with a Max-MEWMA chart proposed by Xie (1999) and the Alternate Variable Multivariate chart by Page (1954). The Max-CUSUM chart had better performance in detecting simultaneous small shifts of process mean and covariance. It also outperformed the other two charts in detecting only mean shifts.

## 4.2 A MCUSUM Chart Based on projection pursuit

Projection pursuit (PP) has been used to develop multivariate control charts for monitoring process mean vector (see for example (Huber 1985; Ngai and Zhang 2001)). Chan and Zhang (2001) applied the PP technique to derive a CUSUM-type chart for variability and called it $MCD_1$.

The PP approach contains two key steps. Firstly, selecting a univariate dispersion chart with charting statistic $Q_i$, Chan and Zhang (2001) chose the CUSUM chart of Johnson (1962). The EWMA chart proposed by Chang and Gan (1995) is another alternative. Secondly, estimating $\lambda_{\max}$ and $\lambda_{\min}$ iteratively over each time period, and calculate the value of charting statistics $Q_i\,(\lambda_{\max})$ and $Q_i\,(\lambda_{\min})$.

They defined $Q_{ik}^+ = \lambda_{ik}^{\max} - (i - k + 1)k^+$ and $Q_{ik}^- = \lambda_{ik}^{\min} - (i - k + 1)k^-$, where $i$ and $k$ indicate the $i$-th and $k$-th observations, $1 \leq k \leq i$; $k^+$ and $k^-$ are the reference values; the $\lambda_{ik}^{\max}$ and $\lambda_{ik}^{\min}$ are respectively the largest and smallest eigenvalues of the matrix $X_i X_i' + X_{i-1} X_{i-1}' + \cdots + X_k X_k'$. The monitoring statistics are defined as

$$Q_i^+ = max\{0, Q_{i1}^+, Q_{i2}^+, \dots, Q_{ii}^+\}$$

$$Q_i^- = min\{0, Q_{i1}^-, Q_{i2}^-, \dots, Q_{ii}^-\}$$

where $Q_0^+ = Q_0^- = 0$.

$MCD_1$ can be used when either the process mean vector and covariance matrix are known or have to be estimated. It can monitor the process mean and covariance simultaneously but cannot distinguish a mean shift from a covariance change. Chan and Zhang (2001) used average run length ($ARL$) and standard deviation of run length ($SDRL$) to evaluate the performance of the $MCD_1$ chart. The simulation showed that it can achieve a small value of $ARL_1$ and $SDRL$, while keeping the value of $ARL_0$ at a specific level. A simple example with $p = 3$ is used to illustrate the performance of the $MCD_1$ chart. The $MCD_1$ chart can also be used when the sample size is greater than 1.

*4.3 A CUSUM Chart for Monitoring the Covariance Matrix of Multivariate Time Series*

Bodnar and Schmid (2017) modified four existing CUSUM charts and one MEWMA chart for monitoring abnormalities in the process covariance matrix of multivariate time series. These four charts are the $MC_1$, $MC_2$ proposed by Pignatiello and Runger (1990), the $MCUSUM$ chart of Crosier (1988), the $PPCUSUM$ chart of Ngai and Zhang (2001), and the MEWMA chart by Śliwa and Schmid (2005). Bodnar and Schmid (2017) modified these charts for original data as well as for residuals from a linear predictor. They compared the modified charts for original data and residuals with the initial $MEWMA$ chart proposed by Śliwa and Schmid (2005). They recommended the modified $MCUSUM$ control charts, since the modified charts had, both for original data and residuals, better performances in detecting almost all the simulated shifts.

## 5. Multivariate EWMA Dispersion Chart

In this section, we review fourteen articles that propose MEWMA-type dispersion charts for individual observations. We grouped these articles into three subsections according to their monitoring statistics. The first category includes the articles that apply the trace or the $L_1$-, $L_2$ or $L_\infty$- norm as the summary statistics of the estimated covariance matrix. In the second category, we discuss the articles that apply the Alt likelihood ratio as the monitoring statistics. The third category includes articles that are based on LASSO techniques. All the charts discussed in this section are designed to detect changes in normally distributed data.

The first EWMA statistic for monitoring the covariance matrix of individual observations was developed by Yeh, Huwang, and Chien-Wei (2005), and it is defined as

$$\boldsymbol{S}_i^{EW} = \lambda \boldsymbol{X}_i \boldsymbol{X}_i' + (1-\lambda) \boldsymbol{S}_{i-1}^{EW}, \tag{3}$$

where $0 < \lambda < 1$ is a smoothing constant. Later in 2007, Huwang, Yeh, and Wu (2007) applied this EWMA statistic to the transformed data (recall Eq. [1]);

$$\boldsymbol{S}_i^{EWMA} = \lambda \boldsymbol{Y}_i \boldsymbol{Y}_i' + (1-\lambda) \boldsymbol{S}_{i-1}^{EWMA}, \tag{4}$$

where $\boldsymbol{S}_0^{EWMA} = \boldsymbol{I}_p$. However, some proposed charts initialized with $\boldsymbol{S}_0^{EWMA} = \boldsymbol{Y}_1 \boldsymbol{Y}_1'$. Most of the reviewed articles on MEWMA-type dispersion charts for individual observations use the EWMA statistic of Eq. [4].

*5.1 MEWMA-type dispersion charts that apply the trace or norm as the summary statistic*

In this subsection, we discuss seven articles that proposed multivariate dispersion charts that apply the trace as the summary statistic of the estimated covariance matrix. We give information about each chart, such as the subgroup size (n), and $p$ simulated in each of the articles in Table 4.

| Authors | Chart name | n | p | Process parameters in Phase I |
|---|---|---|---|---|
| Yeh, Huwang, and Chien-Wei (2005) | Max-MEWMV | n=1 | 2,3 | Known |

| Huwang, Yeh, and Wu (2007) | MEWMS | n=1 | 2,3 | Known |
| :---: | :---: | :---: | :---: | :---: |
| Gunaratne et al. (2017) | MEWMV | n=1 | 4 to 15 | Known |
| Alfaro and Ortega (2019) | | n=1 | 2,3 | Known |
| Memar and Niaki (2009) | $L_1$-norm and an $L_2$-norm of MEWMS MEWMV | n=1 | 2,3 | Known |
| Memar and Niaki (2011) | MEWMS$_{AT}$ | n=1,2,5 | 2,3,4 | Known |
| Tsai et al. (2012) | AEWMS | n=1 | 2,3,5,7,10 | Known |

Table 4: Overview of MEWMA-type dispersion charts that use trace or L-norm as the summary statistic of the estimated covariance matrix.

The first MEWMA chart for monitoring the process covariance matrix for individual observations was proposed by Yeh, Huwang, and Chien-Wei (2005); it is referred to as the MaxMEWMV chart. This chart is based on the deviation of the in-control covariance matrix from the estimated covariance matrix of Eq. [3] as $D_i = S_i^{EW} - I_p$. Next, the authors computed two statistics, $D_{1i}$ and $D_{2i}$, where $D_{1i}$ is the $L_2$-norm of the diagonal elements of $D_i$ and $D_{2i}$ is the $L_2$-norm of the upper triangular off-diagonal elements of $D_i$. $D_{1i}$ is used to detect changes in the variance and $D_{2i}$ to detect changes in the correlation structure. The monitoring statistic for the MaxMEWMV chart is defined as

$$MaxD_i = max\left(\frac{D_{1i} - E(D_{1i})}{\sqrt{Var(D_{1i})}}, \frac{D_{2i} - E(D_{2i})}{\sqrt{Var(D_{2i})}}\right),$$

where the expectation and variance of $D_{1i}$ and $D_{2i}$ are derived analytically. Note that the MaxMEWMV chart is designed with the assumption that $X_i \sim N(\mathbf{0}, I_p)$. However, if the observations are not standard normal, $I_p$ may not shrink $S_i^{EW}$ towards the zero matrix, and thus, $D_i$ will be approximately the same as $S_i^{EW}$. Then, we expect that the chart may have poor performance.

Later in 2007, Huwang, Yeh, and Wu (2007) proposed the first MEWMA chart that applied the trace of the estimated covariance matrix obtained from the EWMA statistic ($S_i^{EWMA}$, Eq. [4]) as the monitoring statistic. This chart is referred to as the MEWMS chart, and the control limits are

$$LCL/UCL = p \pm L\sqrt{2pc_i},$$

where $c_i = \frac{\lambda}{2-\lambda} + \frac{2-2\lambda}{2-\lambda}(1-\lambda)^{2(i-1)}$ and $L$ is the control charting constant.

Also, Huwang, Yeh, and Wu (2007) developed the MEWMV chart to simultaneously monitor shifts in the process mean vector and the covariance matrix. However, the authors showed that the MEWMS chart has better performances than the MEWMV chart in detecting the shifts in the process mean vector and/or process covariance matrix.

In addition, Yeh, Huwang, and Chien-Wei (2005) and Huwang, Yeh, and Wu (2007) discussed multiple CUSUM charts and multiple EWMA dispersion charts. These charts were initially suggested by Hawkins (1991), but the author did not explain them in detail. Yeh, Huwang, and Chien-Wei (2005) and Huwang, Yeh, and Wu (2007) mentioned that multiple CUSUM or EWMA charts are not effective in detecting changes in the correlation.

Applying the trace to detect changes in the process covariance matrix may fail because in several out-of-control situations while some values of the diagonal elements of the estimator increase, some other elements may decrease. Thus, Memar and Niaki (2009) improved the MEWMS and MEWMV charts by applying an L$_1$-norm and an L$_2$-norm to the diagonal elements of $S_i^{EWMA} -$ $I_p$. The authors showed that their proposed charts are quicker in detection than the MEWMS and MEWMV charts except when there is a shift in only the process correlation.

Gunaratne et al. (2017) developed a powerful algorithm based on Parallelized Monte Carlo simulation to enhance the ability of the MEWMS and MEWMV charts in detecting shifts of the process variability for high dimensional data. The authors employed this algorithm to estimate the optimal control limits (L) for the charts for up to $p = 15$. Then the mathematical formulation was developed from the estimated $L$ values, $L = a * e^{b*p} + c * e^{d*p}$, where $a, b, c$, and $d$ are constants and $p$ is the number of correlated quality characteristics. The authors verified and validated the performance of the algorithm and the mathematical formulation with simulation.

In addition, Ajadi and Zwetsloot (2019) argued that the MEWMS chart is ineffective in detecting a decrease in the process variances because the control limits of the chart are designed as symmetric limits; however, nonsymmetric limits are more appropriate.

Memar and Niaki (2011) showed that trace of the $S_i^{EWMA}$, in Eq. [4], can be approximated as a sum of independent chi-squared variables;

$$tr[S_i^{EWMA}] \sim \frac{p}{v}\chi^2(v), \text{ where } v = \frac{np(2-\lambda)}{\lambda}.$$

The authors named this chart, MEWMS$_{AT}$; the control limits are given by $UCL = \frac{p}{v}\chi^2_{\frac{\alpha}{2}}(v)$ and $LCL = \frac{p}{v}\chi^2_{1-\frac{\alpha}{2}}(v)$. Later in 2017, Hwang (2017) proposed Shewhart-type charts inspired on MEWMS$_{AT}$, for details, see Section 6.2.

Tsai et al. (2012) proposed an adaptive time-weighted multivariate EWMA control chart and named it AEWMS chart. The monitoring statistic of this chart is defined as:

$$S_i^{AEWMA'} = \begin{cases} \lambda_1 tr(Y_i Y_i') + (1-\lambda_1) S_{i-1}^{AEWMA'}, & \varepsilon_i < -\gamma \\ \lambda_2 tr(Y_i Y_i') + (1-\lambda_2) S_{i-1}^{AEWMA'}, & |\varepsilon_i| \leq \gamma \\ \lambda_3 tr(Y_i Y_i') + (1-\lambda_3) S_{i-1}^{AEWMA'}, & \varepsilon_i > \gamma \end{cases}$$

where $S_0^{AEWMA'} = tr(Y_1 Y_1')$, $\lambda_1 = \frac{(1+\lambda)}{2}$, $\lambda_2 = \lambda$, $\lambda_3 = \frac{(3+\lambda)}{2}$, $0 < \lambda \leq 1$, $\varepsilon_i = tr(Y_i Y_i') - tr(S_{i-1}^{EWMA})$ (recall $S_i^{EWMA}$ in Eq. [4]), and $\gamma$ is a positive constant. The adaptive adjustment procedure increases the sensitivity of the AEWMS chart for detecting changes in the process variability. Note that $\lambda_3$ gives a value greater than 1 irrespective of the value of $\lambda$. This is a strange value since the smoothing parameter of an EWMA statistic is always between zero and one.

Alfaro and Ortega (2019) mentioned that the quantity $Y_i Y_i'$ used by Huwang, Yeh, and Wu (2007) in Eq. [4] is influenced by the mean. Thus, the authors proposed using the sample covariance matrix of the data set $\{Y_k\}_{k=1,2,\ldots,i}$ to replace the $Y_i Y_i'$. Their proposed chart was shown to be better than the MEWMS chart when the shift in the process variance is very large. However, they only used zero state performance measures, we expect that for steady state performance their results will be less favorable as they incorporate all information from $k = 1$.

## 5.2 MEWMA Charts Based on the Alt Likelihood Ratio

In this subsection, we discuss three articles that proposed multivariate dispersion charts that apply the Alt's likelihood ratio as the monitoring statistic. We give information about each chart, such as the subgroup size (n), and $p$ simulated in each of the articles in Table 5.

| Authors | Chart name | n | p | Process parameters in Phase I |
|---|---|---|---|---|
| Hawkins and Maboudou-tchao (2008) | MEWMC | n=1 | 5,10 | Known |

| Zhang, Li, and Wang (2010) | ELR | n=1,2,4,5 | 2,3,5 | Known |
| Maboudou-Tchao and Hawkins (2011) | SSMEWMAC | n=1 | 5,10,20 | known |

Table 5: Overview of MEWMA-type dispersion charts based on the Alt likelihood ratio statistic.

The multivariate exponentially weighted moving covariance matrix (MEWMC) chart proposed by Hawkins and Maboudou-tchao (2008) is developed by applying the Alt's likelihood ratio statistic to compare $\boldsymbol{S_i^{EWMA}}$ with the identity matrix. The monitoring statistic of the MEWMC chart is defined as

$$c_i = tr(\boldsymbol{S_i^{EWMA}}) - \log|\boldsymbol{S_i^{EWMA}}| - p$$

where $c_i$ is compared to an upper control limit. The authors mentioned that the MEWMC chart can detect both increases and decreases in the process variability. The authors also suggested combining the MEWMC chart with the MEWMA chart that was proposed by Lowry et al. (1992) in order to detect shifts in both process mean vector and covariance matrix.

Zhang, Li, and Wang (2010) developed a chart that integrates the EWMA procedure with the generalized likelihood ratio statistic and named it ELR chart. The authors mentioned that the ELR chart is similar to the MEWMC chart. However, the ELR chart has the advantage of simultaneously detecting shifts in the process mean vector and the covariance matrix. According to the authors, the ELR chart has the benefit of being sensitive to different shifts in the process. In addition, it does not require additional parameters except for the smoothing parameter and control limit. Finally, it can be used for the case when $n \geq 1$. The authors showed that the ELR chart has a better performance than the MEWMC chart only when detecting shifts in the process mean vector.

Maboudou-Tchao and Hawkins (2011) proposed a chart that combines the self-starting multivariate EWMA statistic with a moving covariance matrix (SSMEWMAC). This chart has an advantage over the MEWMC because it does not need a large Phase I sample. The monitoring starts at observation, $i = p + 2$. Maboudou-Tchao and Hawkins (2011) computed the following statistics;

$$\boldsymbol{U_i} = \lambda Y_i + (1 - \lambda)\boldsymbol{U_{i-1}},$$

$$\boldsymbol{V_i} = \lambda Y_i Y_i' + (1 - \lambda)\boldsymbol{V_{i-1}},$$

for $i \geq p + 2$ where $0 < \lambda < 1$, $\boldsymbol{U_{p+1}} = 0$ and $V_{p+1} = I_p$. Then, they computed $M_i = \boldsymbol{U_i'} \Sigma_{U_i}^{-1} \boldsymbol{U_i}$ and $\widetilde{c}_i = tr(\boldsymbol{V_i}) - \log|\boldsymbol{V_i}| - p$ to monitor shifts in the process mean vector and variability, respectively. Both statistics ($M_i$ and $\widetilde{c}_i$) are compared against control limits and a signal is received if one of them exceeds their limit. The authors mentioned that self-starting methods cannot detect shifts when the process deviate immediately from the in-control process after the monitoring begins.

*5.3 MEWMA-type Dispersion Charts Based on LASSO Method*

In this subsection, we discuss multivariate control charts that apply the LASSO method in building up the statistics for monitoring changes of the process variability for details on LASSO see for example, (d'Aspremont, Banerjee, and El Ghaoui 2008; Friedman, Hastie, and Tibshirani 2008; Rothman et al. 2008). We give information about each chart, such as the subgroup size (n), and $p$ simulated in each of the articles in Table 6.

| Authors | Chart name | n | p | Process parameters in Phase I |
|---|---|---|---|---|
| Yeh, Li, and Wang (2012) | LASSO-MEWMC | n=1 | 5,10,20 | Known |
| Maboudou-Tchao and Diawara (2013) | S-LEWMC | n=1 | 5,10,20 | Known |
| Li, Wang, and Yeh (2013) | PLR | n=50 | 5,10 | Known |
| Wang, Yeh, and Li (2014) | pGLR pMEWMAC | n=1 | 5 | Known |
| Shen, Tsung, and Zou (2014) | MaxNorm | n=1,20,50 | 5,10,30 | Known |
| Fan et al. (2017) | EIGEN | n=1 | 5,10 | Known |

Table 6: Overview of MEWMA-type dispersion charts based on LASSO statistics.

In 2009, Zou and Qiu. (2009) introduced the LASSO technique to Statistical Process Control (SPC) for detecting shifts in a few components of the process mean vector. Their proposed chart is also effective as a post signal diagnostic tool for identifying the shifted variable(s).

The LASSO method has also been proposed for detecting shifts that occur in only few elements of the covariance matrix. For instance, Yeh, Li, and Wang (2012) improved the MEWMC chart (recall Section 5.2) with the LASSO technique and named it the LASSO-MEWMC chart. The authors defined $\widehat{\Omega}_i$ as the solution for the penalized likelihood function

$$\widehat{\Omega}_i = \arg\min_{\Omega_i > 0}\{tr(\Omega_i S_i^{EWMA}) - \log|\Omega_i| + \rho \parallel \Omega_i - I_p \parallel_1\}$$

where $\Omega_i$ is the precision matrix of the process, and $\rho > 0$ is a tuning parameter to achieve different sparsity levels for the $\Omega_i$ estimate. Note that $\widehat{\Omega}_i = \widehat{\Sigma}^{-1}$, where $\widehat{\Sigma}$ is an estimate of the covariance matrix of the process. The penalty function used by Yeh, Li, and Wang (2012) shrinks the estimate of the precision matrix towards the in-control covariance matrix, $I_p$. Thus, the monitoring statistic of the LASSO-MEWMC chart is defined as

$$l_i = \log|\widehat{\Omega}_i| - tr\left(\widehat{\Omega}_i S_i^{EWMA}\right) + tr(S_i^{EWMA}),$$

Maboudou-Tchao and Diawara (2013) also proposed a LASSO chart for monitoring the covariance matrix (S-LEMWC). The setup of their proposed chart is slightly different from that of LASSO-MEWMC. In S-LEMWC chart, the graphical LASSO algorithm developed by Friedman, Hastie, and Tibshirani (2008) was used to obtain the solution $\widehat{\Omega}_i$

$$\widehat{\Omega}_i = \arg\max_{\Omega>0}\{tr(\Omega_i Y_i Y_i') - \log|\Omega_i| + \rho \parallel \Omega_i \parallel_1\}. \tag{5}$$

Then the authors computed the EWMA statistic

$$\overline{\overline{S}}_i = \lambda \widehat{\Omega}_i^{-1} + (1-\lambda)\overline{\overline{S}}_{i-1}.$$

The S-LEWMC chart is the same as the MEWMC chart when $\rho = 0$ ($i.e$ no regularization is applied).

Also in 2013, Maboudou-Tchao and Agboto (2013) modified the penalized likelihood ratio (PLR) chart proposed by Li, Wang, and Yeh (2013) such that it can monitor the process variability of data with a sample size less than $p$. The authors named the chart LASSO. The LASSO chart is also based on using a graphical LASSO algorithm proposed by Friedman, Hastie, and Tibshirani (2008). The penalty function used by both LASSO and S-LEWMC charts in Eq. [5] can shrink the sample precision matrix towards the zero matrix. However, the penalty function used by Yeh, Li, and Wang (2012) is more realistic since it penalizes towards the in-control covariance matrix.

In addition, Wang, Yeh, and Li (2014) developed two penalized likelihood estimation control charts that can simultaneously monitor the process mean vector and covariance matrix. The first chart is the penalized generalized likelihood ratio (pGLR); the second chart is the penalized versions that combines the MEWMA and MEWMC charts (pMEWMAC). In these charts, the authors defined the $\hat{\mu}_i$ and $\hat{S}_i^\lambda$ as the solution for the penalized likelihood function for the process mean vector and covariance matrix, respectively. These are provided as follows

$$\hat{\mu}_i = \arg\min_{\mu}\{(E_i - \mu)' \Sigma_0^{-1}(E_i - \mu) + \rho_1 \sum_{k=1}^{p} |\mu_k|\},$$

$$\hat{S}_i^\lambda = \arg\max_{S>0}\{-\ln|S| - tr(S^{-1}S_i^{EWMA}) + \rho_2 \parallel S - \Sigma_0 \parallel_1\}$$

where $\mu_k$ is the $k^{th}$ element of vector $\mu$. Also, the $\rho_1$ and $\rho_2$ are the penalty coefficients for the process mean vector and covariance matrix respectively. $E_i$ is the MEWMA statistic for monitoring the process mean vector and $S$ is the estimated covariance matrix.

In addition, Shen, Tsung, and Zou (2014) introduced a MaxNorm chart. This chart is also effective in detecting the shifts that occur in only a few elements of the covariance matrix. The performance of the penalized likelihood control charts depends on the choice of the regularization parameter. However, the MaxNorm chart avoids setting the regularization parameter.

The setup of MaxNorm chart is similar to the MaxMEWMV chart by Yeh, Huwang, and Chien-Wei (2005). It is based on deviating the in-control covariance matrix from the estimated covariance matrix of Eq. [4] as $T_i = S_i^{EWMA} - I_p$. Then the authors computed two statistics, $T_{1i} = \sum_{k=1}^{p}\sum_{l=1}^{p} c_{i(k,l)}^2$ and $T_{2i} = max(|c_{i(1,1)}|, \cdots, |c_{i(p,p)}|)$, where $c_{i(k,l)}$ is the combination of the diagonal elements and the upper triangular off-diagonal elements of $T_i$ (i.e. $c_{i(k,l)}$ for $k \leq l$). The $T_{1i}$ and $T_{2i}$ statistics are respectively the $L_2$-norm and $L_\infty$−norm of $c_{i(k,l)}$. The monitoring statistic is defined as

$$MaxNormT_i = max\left(\frac{T_{1i}-E(T_{1i})}{\sqrt{Var(T_{1i})}}, \frac{T_{2i}-E(T_{2i})}{\sqrt{Var(T_{2i})}}\right),$$

where the expectation and variance of $T_{1i}$ and $T_{2i}$ statistics are obtained through simulation.

Later in 2017, Fan et al. (2017) proposed a chart that uses the eigenvalues of $S_i^{EWMA} - I_p$ to monitor changes in the process covariance matrix and named it as the EIGEN chart. The set-up of the EIGEN chart is similar to that of MaxNorm chart. Like the MaxNorm chart, the EIGEN chart is also efficient in detecting changes that occur in only a few elements of the covariance matrix without using any tuning parameter to achieve sparsity of the precision matrix.

## 6. Shewhart-type Dispersion Chart

In this section, we discuss 6 articles on Shewhart-type dispersion charts for monitoring individual observations. All the charts are designed for detecting changes in normally distributed data. Mason, Chou, and Young (2009), Mason, Chou, and Young (2010), and Djauhari (2010) used Wilks' statistic to build charting statistics. The charts proposed by Maboudou-Tchao and Agboto (2013), and Hwang (2017) are based on the trace of the covariance matrix. Both Hwang (2017) and Khoo and Quah (2003) used Chi-squared distribution to approximate the monitoring statistics. We will briefly introduce these Shewhart-based methods, we discuss the development of these methods and their internal relations. Their performances, if available, will also be discussed. In Table 7, we give information about each chart, such as the subgroup size (n), and $p$ simulated in each of the articles.

| Articles | Chart name | subgroup size | p | Process parameters in Phase I |
|---|---|---|---|---|
| Khoo and Quah (2003) | Chi-square | n=1 | 2 | Known |
| Mason, Chou, and Young (2009) |  | n=1,2,3,4,5 | 4,6,7 | Estimated |
| Mason, Chou, and Young (2010) | W | NIL | NIL | Estimated |

| Djauhari (2010) | F | NIL | NIL | Estimated |
| Maboudou-Tchao and Agboto (2013) | LASSO | n=1,2.3,4,7 | 5,10 | Known |
| Hwang (2017) | MTSSD MMSSD | n=1 | 2,4 | Known |

Table 7: Overview of Shewhart-type dispersion charts.

## 6.1 *Wilks' Statistic-based Charts*

Wilks' statistic (Wilks (1962, 1963)) is defined as the ratio of the determinants of two scatter matrices;

$$W = \frac{|SS_m|}{|SS_{m+1}|} = \left(\frac{m-1}{m}\right)^p \frac{|S_m|}{|S_{m+1}|} = 1 - \frac{m+1}{m^2}T^2. \qquad (6)$$

Where $SS_m$ is the scatter matrix of the historical data set from Phase I, and $SS_{m+1}$ is the scatter matrix of an augmented dataset which includes both the historical data in Phase I and the newest observation in Phase II. They are calculated as:

$$SS_m = \sum_{j=1}^{m}(X_j - \bar{X}_m)(X_j - \bar{X}_m)' = (m-1)S_m$$

$$SS_{m+1} = \sum_{j=1}^{m}(X_j - \bar{X}_{m+1})(X_j - \bar{X}_{m+1})' + (X_i - \bar{X}_{m+1})(X_i - \bar{X}_{m+1})' = m\,S_{m+1}$$

Furthermore $S_m$ and $S_{m+1}$ are estimators of the covariance matrix based on $m$ and $m+1$ observations, respectively.

In Eq. [6], it is shown that $W$ is related to a Hotelling's $T^2$ statistic. Mason, Chou, and Young (2009) treated $W$ as the charting statistic to design a dispersion chart and used a $Beta$ distribution to define the control limits. One property of Wilks' statistic, is that its value is between 0 and 1.

Since the determinant of $SS_{m+1}$ is always greater than the determinant of $SS_m$, a smaller value of the ratio of the determinants indicates a greater difference between $SS_m$ and $SS_{m+1}$ (Wilks (1962, 1963)). Values of $W$ near 0 indicate that the estimated covariance matrix $S_{m+1}$ is different from $S_m$, and that there is a possible shift in process variation. Therefore, a $W$ chart with lower bound is enough to detect changes in $S_{m+1}$ compared with $S_m$ (Mason, Chou, and Young (2009)).

Diagnosing the cause of an out-of-control signal is a challenge in monitoring multivariate process variability. To solve this, Mason, Chou, and Young (2010), proposed a technique to decompose Wilks' statistic and identify the process variables contributing to the signal.

Djauhari (2010) derived an $F$ chart from $SS_m$ and $SS_{m+1}$ as a complement to Wilks' statistic, and suggested to run the $W$ chart and $F$ chart simultaneously. From their comparison results, it followed that the $W$ chart outperforms the $F$ chart in detecting correlation shifts. The $F$ chart is more sensitive to detect variance changes.

The $W$ chart can be applied when the sample size is greater than 1 and less than $p$ (Mason, Chou, and Young (2009, 2010)). The monitoring procedure proposed by Djauhari (2010) is only suitable for individual observations.

## 6.2 Trace-based Charts

The $F$ statistic in Djauhari (2010) is equal to the square root of the trace of $(SS_{m+1} - SS_m)^2$. The trace of the covariance matrix represents the total variation, some authors use this metric to construct charting statistics for process monitoring. Hotelling (1947) firstly used trace to monitor $\Sigma$ with a Shewhart chart. Maboudou-Tchao and Agboto (2013) used the trace combined with the log determinant of the estimated covariance matrix as the monitoring statistic

$$L_i = tr\left(\tilde{S}_i\right) - \log\left|\tilde{S}_i\right| - p.$$

Here $\tilde{S}_i$ is the graphical LASSO estimator of the covariance matrix, for more details about LASSO see section 5.3. $\tilde{S}_i$ can also be explained as a penalized likelihood estimator of the covariance matrix.

Maboudou-Tchao and Agboto (2013) compared the proposed LASSO chart with Alt's chart proposed by Alt (1985), the conditional entropy (CE) chart Guerrero-Cusumano (1995), and the Exponentially Weighted Moving Covariance Matrix (MEWMC) chart (see Section 5.2), and used the ARL to evaluate and compare their performances. The LASSO chart outperforms Alt's chart and the CE chart in detection of variance shifts. It has worse performance under correlation shifts compared to the MEWMC chart and Alt's chart. It can detect simultaneous variance and correlation shift quickly. Maboudou-Tchao and Agboto (2013) concluded that the LASSO chart is workable for individual observations, but it has better performance with grouped data in phase II monitoring.

Hwang (2017) applied the trace to the covariance estimator $S_m$ to represent the total variance of the $p$ variables. He standardized the observations as Eq. [1], and calculated the monitoring statistic as $tr(S_i)$, where $S_i = \sum_{k=1}^{i} Y_k Y_k'$, $k = 1,2 \dots i$. This chart is called the multivariate trace sum of squared deviation (MTSSD) chart. It can detect variation shifts. Hwang (2017) also proposed the multivariate matrix sum squared deviation (MMSSD) chart. The MMSSD chart is not based on the trace concept, but on the sum of all elements of $S_i$, so it can monitor the total variation and correlation simultaneously. The control limits for the MTSSD and MMSSD charts are obtained from two Chi-square distributions separately.

Hwang (2017) measured the performances of charts by evaluating the median run length (MRL) and standard deviation run length (SDRL). The comparison included the MEWMS$_{AT}$ and the MEWMSL$_1$ charts, for a detailed introduction see Section 5.1. The simulation for individual observations revealed that, the MEWMS$_{AT}$ and MEWMSL$_1$ charts are incomparable because of the extremely small in-control MRL and very large SDRL. The MTSSD chart was effective for small variance shifts, and MMSSD can detect correlation shifts well. Both of them can adapt to the case where the subgroup size is greater than 1.

Khoo and Quah (2003) designed another multivariate chart for monitoring process dispersion based on individual observations. Their monitoring statistic is the successive differences between the multivariate observations and can be expressed as $K_i = \frac{1}{2}(X_i - X_{i-1})'\Sigma_0^{-1}(X_i - X_{i-1})$. Khoo and Quah (2003) also used a Chi-squared distribution to set the control limits. The chart may not be reliable because $K_i$ is affected by serial correlation and the control limits are derived under the independence assumption.

## 7. Non-Parametric Dispersion Charts

All previous discussed control charts are based on the assumption that the data can be modeled by a multivariate normal distribution, but in many applications, the data come from unknown or nonnormal distributions. Then, conventional control charts are unreliable, especially when the sample size is small (Li et al., 2013). Nonparametric control charts, which are more robust to the underlying data distributions, are useful for monitoring non-normal distributed data. Chakraborti and Graham (2019) provided an excellent review of nonparametric (distribution-free) control charts. According to this review paper, distribution-free statistics based on signs and ranks are widely used to build monitoring statistics. For details about spatial signs and rank, see Sirkiä et al. (2009). Here we focus on nonparametric dispersion charts for multivariate individual observations. Two charts have been introduced in the literature. Li et al. (2013) proposed their control chart based on a spatial sign covariance matrix, Huwang, Lin, and Yu (2019) used spatial rank to build charting statistics and compared this chart with the chart proposed by Li et al. (2013). In addition, one of the articles( Huwang et al. (2017)) in Section 8 for high dimensions is also distribution free, but it is excluded in this section for the sake of repetition. Table 8 provides the information about each chart, such as the subgroup size (n), and $p$ simulated in each of the two articles.

| Authors | Chart name | n | p | Process parameters in Phase I |
|---|---|---|---|---|
| Li et al. (2013) | MNSE | $n = 1$ | 2,3,4,5,7, 10,15,20,30 | Estimated |
| Huwang et al. (2017) | Dissimilarity | n=1,6,11,21 | 5,10,20 | Estimated |
| Huwang, Lin, and Yu (2019) | MSRE | $n = 1$ | 3,5 | Estimated |

Table 8: Overview of nonparametric charts for monitoring shape parameter.

### 7.1 A Nonparametric EWMA Chart Based on Spatial Sign

Li et al. (2013) used a multivariate spatial sign test and exponentially weighted moving average (EWMA) scheme to develop a distribution-free control chart for monitoring shape parameters of the underlying data.

For independent and identically distributed (i.i.d.) historical $p$ - dimensional observations $X_j, j = -m + 1 \ldots - m + j, \ldots 0$, the spatial sign function is defined as

$$U(X) = \begin{cases} \|X\|^{-1}X, & X \neq 0 \\ 0, & X = 0 \end{cases}$$

where $\|X\| = (X'X)^{1/2}$.

Next, the multivariate affine equivariant median vector $\theta_0$ and the associated transformation matrix $A_0$ need to be estimated from historical data in Phase I by solving

$$\frac{1}{m} \sum_{j=-m+1}^{0} U\left(A_0(X_j - \theta_0)\right) = 0$$

$$\frac{1}{m} \sum_{j=-m+1}^{0} U\left(A_0(X_j - \theta_0)\right) U'\left(A_0(X_j - \theta_0)\right) = \frac{I_p}{p}.$$

The roles of the affine equivariant median $\theta_0$ and the transformation matrix $A_0$ are similar to those of the sample mean and covariance matrix, in constructing conventional control charts ( Huwang, Lin, and Yu (2019)).

On-line observations $X_i, i = 1, 2, ...$, can be standardized and transformed to a unit vector $v_i$ through $v_i = U(A_0(X_i - \theta_0))$. The EWMA statistic can be computed as

$$w_i = (1 - \lambda)w_{i-1} + \lambda v_i v_i'$$

where the initial matrix $w_0$ is defined as $I_p/p$. Finally, the charting statistic is defined as

$$Q_i = \sqrt{\frac{2 - \lambda}{\lambda} \cdot tr((pw_i - I_p)^2)}$$

Li et al. (2013) named this chart the multivariate nonparametric shape EWMA (MNSE) chart. They compared the MNSE chart with the MEWMC chart proposed by Hawkins and Maboudou-tchao (2008) (see Section 5.2) and evaluated the steady-state ARL behavior of each chart. They considered one normal and five non-normal distributions under different scenarios. The results showed that, the MNSE chart was robust to non-normal distributions and efficient in detecting small and moderate shifts ( Li et al. (2013)). It was also effective in detecting downward shifts, but its performances was not as good as the MEWMC chart for large shifts under normal and nonnormal distributions.

The MNSE chart also performed well in the real-data example, from a white-wine production process. The MNSE chart can be adapted to the case of subgroups with more than one observation.

*7.2 A Nonparametric EWMA Chart Based on Spatial Rank*

Huwang, Lin, and Yu (2019) proposed a nonparametric multivariate EWMA chart for monitoring shape matrices. This chart is based on a spatial rank test proposed by Sirkiä et al. (2009), which used the estimated spatial rank covariance matrix for monitoring the shape matrix of a multivariate process.

For Phase II observations, $X_i, i = 1, 2, 3, \ldots$, the empirical spatial rank function, which is the average of the spatial signs of pairwise differences for incoming observations can be computed as

$$R_E^*(X_i) = \frac{1}{m} \sum_{j=-m+1}^{0} U\left(M_0(X_i - X_j)\right)$$

where $(M_0'M_0)^{-1} = \frac{1}{m-1}\sum_{j=-m+1}^{0}(X_j - \bar{X})(X_j - \bar{X})'$, $\bar{X}$ is the mean vector of $m$ historical observations. Then, an EWMA covariance matrix estimate based on the empirical spatial rank function $R_E^*(X_i)$ is defined as

$$S_i^{R_E} = \lambda R_E^*(X_i)R_E^*(X_i)' + (1-\lambda)S_{i-1}^{R_E},$$

where $S_0^{R_E} = \frac{1}{m}\sum_{j=-m+1}^{0} R_E^*(X_j)R_E^*(X_j)'$. The monitoring statistic of this chart is defined as

$$Rank_i = \left[C_p \boldsymbol{vec}(S_i^{R_E})\right]' \left\{Cov[C_p \boldsymbol{vec}(S_i^{R_E})]\right\}^{-1} \left[C_p \boldsymbol{vec}(S_i^{R_E})\right].$$

Where $C_p$ is the projection matrix, $\boldsymbol{vec}(S_i^{R_E})$ is the vectorization of $S_i^{R_E}$ obtained by stacking the columns on top of each other. Since $C_p$ is rather complex, for detailed explanation and derivation see Huwang, Lin, and Yu (2019). The proposed chart is called multivariate spatial rank EWMA (MSRE) chart.

Huwang, Lin, and Yu (2019) compared the performance of the MSRE chart with the MNSE chart proposed by Li et al. (2013) (see Section 7.1). The results showed that, under different distributions, the MSRE chart outperformed MNSE chart in detecting increasing variance shifts especially when the magnitudes of the shifts are large. The MSRE chart had better performances when both variance and correlation shifts occur simultaneously. For other type of changes, the performances of MSRE chart were unstable for different distributions. Huwang, Lin, and Yu (2019) used the same white-wine production data with Li et al. (2013) to illustrate the applicability of the proposed MSRE chart.

Since the MSRE chart is not as distribution-free as MNSE chart, it is necessary to use the resampling method to obtain the desired upper control limit Huwang, Lin, and Yu (2019). Another weakness of both two charts is that a large amount of data are required to reflect the in-control distribution in Phase I.

## 8. Multivariate Dispersion Charts for High Dimensional Data

In this section, we discuss three papers focused on multivariate dispersion charts that are designed for high dimensional data. We provide the information about each chart, such as the subgroup size (n), and $p$ simulated in each of the articles in Table 9.

| Authors | Chart | n | p | Process parameters in Phase I |
|---|---|---|---|---|

| Huwang et al. (2017) | Dissimilarity | n=1,6,11,21 | 5,10,20 | Estimated |
| Li and Tsung (2019) | MVP | n=1 | 5,10,20,30 | Known |
| Kim et al. (2019) | RPLR | n=5,10 | 10 | Known |

Table 9: Overview of the dispersion charts for monitoring high dimensional data.

With high-dimensional data, we mean applications with large $p$, $i.e.$ with many variables. There is no clear cut-off point between "small $p$" and "large $p$". Here we include those articles that classify themselves as monitoring tools for high-dimensional data. Shu and Fan (2018) mentioned that the distribution of high dimensional data is in most cases unknown and non-normal, and thus nonparametric control charts are useful in these situations. Apart from the three discussed articles in this section, already discussed methods like Section 5.3, and the non-parametric methods in Section 7 can also be used for high-dimensional data.

Huwang et al. (2017) proposed a chart that applies an EWMA statistic for monitoring the covariance matrix based on the idea of a dissimilarity index. The dissimilarity index method was initially introduced by Kano et al. (2002) for comparing distributions of two data sets to detect changes in operating conditions of chemical processes.

To set up this method, the authors first computed the EWMA statistic as

$$\widehat{M}_i = (1 - \lambda)\widehat{M}_{i-1} + \lambda X_i X_i'$$

where $0 < \lambda \leq 1$, and $\widehat{M}_0 = \hat{\Sigma}_0$. Note that it is assumed that $\hat{\Sigma}_0$ can be obtained from Phase I. Then, they provided their proposed dissimilarity index as

$$D_i = \frac{4}{p} tr(B_i^2 - B_i) + 1$$

where $B_i = (\hat{\Sigma}_0 + \widehat{M}_i)^{-1} \widehat{M}_i$ and a signal is received whenever $D_i > UCL$.

The index $D_i$ measures the dissimilarity between the EWMA covariance matrix $\widehat{M}_i$ and the estimated covariance matrix $\widehat{\Sigma}_0$. Their proposed chart was shown to have better performance than MEWMC in most cases. The authors also mentioned that the chart is a non-parametric chart.

Also, Li and Tsung (2019) proposed a MEWMV-based chart for monitoring the process variability (the MVP chart). The MVP chart combines a powerful high-dimensional covariance matrix test with EWMA statistics to monitor high dimensional variability based on individual observations. This chart is also efficient in detecting changes that occur in only a few elements of the covariance matrix without using any tuning parameter.

In addition, the MVP control chart was shown to be more effective than the MaxNorm, LASSO-MEWMC and, PLR charts when there is a shift in the process variance but less effective for detecting a shift in the covariance matrix when $p$ is not large. Thus, the superiority of the MVP chart to its counterparts increases as the value of $p$ increases.

Also, Kim et al. (2019) proposed another multivariate dispersion control chart for high dimensional data. This chart is based on ridge penalized likelihood ratio (RPLR). The performance of the chart was shown to be better than the PLR control chart (recall Section 5.3).

## 9. Conclusion and Future Research Ideas

Multivariate dispersion charts are used to detect process dispersion shifts. Often in applications, only limited or individual observations are available. Therefore, conventional dispersion charts, which are designed for grouped data, cannot provide satisfactory solutions. Various dispersion charts are proposed for monitoring individual observations. The objective of our paper is to provide an overview of the existing multivariate dispersion charts for individual observations. We have briefly introduced their underlying theories and monitoring schemes. We also included the performance comparisons that are available in the reviewed literatures.

We included 30 papers and classified them based on the type of chart they proposed. The reviewed Shewhart type charts, CUSUM type charts, and EWMA type charts are all designed for monitoring data following a multivariate normal distribution. Various techniques are used to build charting statistics, some are dominant in a specific type of chart. For example, the Wilks' statistic are used in three Shewhart charts, Alt's likelihood ratio and trace methods are popular in EWMA charts. Two nonparametric charts are also reviewed since they are quite useful when the distribution of the data is unknown. The monitoring statistics for nonparametric charts are derived from spatial sign and rank statistics.

CUSUM charts are effective in monitoring small and persistent process changes and were reviewed in Section 4. Both the Max-CUSUM chart and $MCD_1$ chart can monitor the process mean and variability simultaneously. The $MCD_1$ chart and the chart proposed by Healy (1987) can adapt to the cases where the sample size is greater than 1. Since there are only four CUSUM based dispersion charts, more methods can be explored in the future.

In addition, we reviewed 14 articles that discuss the multivariate EWMA control chart for individual observations. We grouped them into 3 sections which include the charts that apply the trace, the Alt likelihood ratio as the monitoring statistic as well as the charts that are based on LASSO techniques.

Six Shewhart type charts are discussed in Section 6. We find that Wilks' statistic is useful in building charting statistics (see Table 7). Trace is also a valid technique for designing Shewhart charts (Maboudou-Tchao and Agboto (2013); Hwang (2017)). But all the Shewhart based charts are for monitoring process variability only. A possible extension of these methods is modifying them to monitor both location and scale parameters simultaneously.

Nonparametric charts are effective on monitoring data from unknown distributions. Spatial sign and rank tests are useful techniques for designing distribution-free charts. We introduced MNSE and MSRE charts in Section 7 and discussed their performances in different scenarios. Both of them focused on Phase II monitoring and have the same drawback, that they require a large amount of data to estimate the in-control distribution and decide the control limits in Phase I. Li et al. (2013) discussed three future research directions based on the MNSE chart: 1) nonparametric charts than can both monitor the location and covariance structure are needed; 2) a self-starting version of the MNSE charts is worth to explore; 3) extend the MNSE charts to Phase I analysis. We totally agree with these recommendations and think they can also be applied to MSRE chart proposed by Huwang, Lin, and Yu (2019).

In addition, we identify the following research directions.

(i) With the development of data collection techniques, more variables are becoming accessible. More multivariate dispersion control charts for high dimensional data need to be developed.

(ii) Some methods only monitor the variances, for example by looking at the trace, while other methods monitor the entire variance covariance matrix by, for example, a monitoring statistic based on the determinant. It is unclear which method is preferred for what application.

(iii) Some methods have incorrect control limits. As the limits are designed as symmetric limits, whereas the monitoring statistics follow skewed distributions. More reliable algorithm or simulation methods for deciding correct control limits need to be proposed.

(iv) Most of the methods are ineffective in detecting decreases in the process variances.

(v) Also, we find that there is no chart designed for monitoring transient shifts, which is worth to be considered.

(vi) Steady state performance should be evaluated. Some methods work well for zero-state performance but may have poor steady state performance.

(vii) Robustness of methods to outliers and non-normalities should be investigated.

(viii) In most of the articles, the process parameters are assumed known. In practice, parameters are estimated, therefore, the effects of parameter estimation should be investigated.

(ix) We observed that less research has been done on non-parametric, MCUSUM and high dimensional data charts, more work is encouraged in this area.

(x) Finally, all the charts we reviewed are designed for monitoring independent observations, if this assumption is invalid, we need new charts to deal with this situation.